\documentclass[preprint,aps,prb,citeautoscript,showpacs]{revtex4} 
\usepackage{graphicx} 
\usepackage{amsmath} 
\usepackage{psfrag} 
\begin{document} 
 
\title{Confined free-electrons in an applied electric field:\\ Discontinuous electron density} 
 
\author{St\'ephane Pleutin} 
\affiliation{Physikalisches Institut, Albert-Ludwigs-Universit\"at, 
Hermann-Herder-Stra{\ss}e 3, D-79104 Freiburg, Germany}

\date{\today} 
\pacs{73.22.-f, 85.35.Be, 03.67.Lx} 
\begin{abstract} 
We consider free electrons in rectangular quantum dots, with 
either hard wall boundary conditions or anharmonic confinement. 
In both cases, due to finite size effects, a homogeneous electric 
field applied along one of the rectangular axis is shown to 
induce abrupt changes in the electron density, parallel and 
perpendicularly to the field direction: the electron density jumps 
from one configuration to another. Making use of this property, we 
propose a purely electrical mechanism to control the magnitude of 
the effective exchange coupling between two quantum dots. This 
system has been proposed recently as a quantum gate for quantum 
computation. 
\end{abstract} 
\maketitle 
 
\section{introduction} 
Due to remarkable advances in microfabrication techniques, it is 
possible nowadays to study the spectroscopic properties of single 
systems at mesoscopic or even nanoscopic scale. These can be 
semiconductor quantum dots \cite{kouwenhoven,reimann}, ultrasmall 
metallic particles \cite{vondelft} or, organic molecules 
\cite{nitzan}, for instance. This opens many new perspectives for 
fundamental physics but also for the development of new 
technologies. Indeed, on the one hand, it is possible to study 
properties of very small systems where quantum phenomena are 
prominent; on the other hand, one may think of applying these 
new properties to go one step further on the way to extreme 
miniaturization towards electronics at the nanometer scale. In 
this context, to understand the behavior of small systems under 
the influence of external excitations is a central issue, with the 
aim to control the parameters of future nanodevices by external 
sources. Here we focus our attention to DC homogeneous electric 
fields applied to semiconductor quantum dots. Since in all the 
above mentioned systems electrons are confined in a very small 
region of space giving rise to a discrete electronic spectrum, we 
expect they should share common features, despite important 
differences. Therefore, we believe some of the conclusions 
described below are relevant for metallic particles and large 
organic molecules, as well. 
 
The problem of the influence of a DC electric field on electronic 
systems has attracted a lot of interest since many years. In any case, 
we don't want to review this tremendous amount of articles but, 
instead, we content ourselves to cite those of direct use for us. 
Most often in the literature, this problem is treated in 
one-dimension, either for free electrons \cite{landau,rabinovitch} or 
for electrons in a periodic potential \cite{rabinovitch,nenciu}, with 
the main focus on the Wannier-Stark ladder problem in the latter 
case. Here we investigate systems with many channels. More 
precisely, we don't consider Zener tunneling between bands, but 
we are rather interested in effects caused by the interplay of the 
applied electric field that distorts the electronic bands and the 
Pauli principle which acts as a kind of quantum pressure. 
 
We consider rectangular semiconductor quantum dots with 
characteristic sizes such that their mean level spacing ranges in 
the meV region. The homogeneous electric field is applied parallel 
to the long axis. The electrons within the box are supposed to be 
non-interacting; in particular, the electric field is assumed to 
be unscreened.  This is a good approximation in the case of 
strictly one dimensional quantum dots, as it was shown recently 
for molecular wires \cite{pleutin1}. It should also be valid for 
quasi-one dimensional  dots where only few channels are occupied 
which is precisely the case of interest in the present work. A 
first obvious effect of the field is to polarize the electronic 
cloud in the longitudinal direction i.e. parallel to the field 
direction. Our purpose is to investigate another effect caused by 
the finite size of the system. We consider two types of 
confinement: infinite potential wall and harmonic confinement 
plus a small anharmonic perturbation. In both cases, the electric 
field is shown to induce at a critical value, $E_c$, abrupt 
changes in the electron density in the two directions, parallel 
and perpendicular to the field. Despite the fact that no interband 
coupling exists, the electric field modifies the distribution of 
the electrons among the different bands. The net effect of this 
redistribution is to decrease the longitudinal polarization and 
to push the electrons toward the surfaces of the well 
perpendicularly to the field. This effect is discussed here on 
the basis of simple models, with the purpose of pointing out 
the basic mechanism at its origin and its possible relevance in 
the context of nanophysics and quantum computation. Indeed, as a 
first application, we consider two lateral weakly coupled quantum 
dots: the electric field is shown to modify abruptly the gap 
between the lowest singlet and triplet states. This last effect 
could have important application in the context of quantum 
computation were similar devices where proposed as possible 
realization of quantum gates \cite{loss}. 
 
The paper is organized as follows. In section II we investigate 
the case with hard wall boundary conditions, and in section III 
the case with harmonic confinement plus a perturbative anharmonic 
term. Finally, in section IV, we present our results for two 
weakly coupled quantum dots before closing with some conclusions. 
 
\section{Rectangular potential well with infinite walls in an applied electric field} 
We start by considering a system of free electrons confined in a rectangular potential well with infinite walls. The corresponding one-electron time-independent Schr\"odinger equation in presence of an applied electric field is 
\begin{equation} 
-\frac{\hbar^2}{2m}\nabla^2\Psi(x,y)+q E x\Psi(x,y)=\epsilon \Psi(x,y) 
\end{equation}where $m$ is the effective mass of the quasi-electrons, $q$ the electron charge ($q<0$) and $E$ the applied electric field. The wave functions fulfill the hard-wall boundary conditions: $\Psi(0,y)=\Psi(L_x,y)=0$ and $\Psi(x,0)=\Psi(x,L_y)=0$. We assume a rectangular box such that $L_x>L_y$ (cf. 
Fig.~1). 
 
Without electric field ($E=0$), we recover one of the basic problems of quantum mechanics: the well known particle in the box problem in two dimensions. The eigenenergies are given by \cite{cohen} 
 
\begin{equation} 
\epsilon_{n_x,n_y}=\frac{\hbar^2}{2m}\big[ (\frac{\pi 
n_x}{L_x})^2+(\frac{\pi n_y}{L_y})^2\big] 
\end{equation}where $\{n_x,n_y\}\in \mathbf{N}^*\times \mathbf{N}^*$ ($\mathbf{N}^*$ is the set of integers without 0). 
 
With electric field ($E \ne 0$), since there is still no coupling 
between the $x$ and $y$ coordinates, the wave function may again 
be written as a product: $\Psi(x,y)=\varphi(x)\xi(y)$. The 
equation for $\xi(y)$ is the one of a free-particle in a one 
dimensional box \cite{cohen}. The equation for $\varphi(x)$ reads 
 
\begin{equation} 
\label{airyequation} 
-\frac{\hbar^2}{2m}\frac{d^2}{dx^2}\varphi(x)+q E x\varphi(x)=e \varphi(x). 
\end{equation}Solutions of this problem are well known for both infinite \cite{landau} and finite size systems \cite{rabinovitch}. We closely follow these two references. 
 
We set ${\tilde e}=\frac{2m}{\hbar^2}e$ and 
$F=-\frac{2m}{\hbar^2}q E$ and write down the solution of 
Eq.~(\ref{airyequation}) as a linear combination of Airy 
functions \cite{abramovitch}, $\mathbf{Ai}$ and $\mathbf{Bi}$, 
\begin{equation} 
\varphi(x)=\alpha \mathbf{Ai}\big(-F^{1/3}[x+\frac{{\tilde e}}{F}]\big)+\beta \mathbf{Bi}\big(-F^{1/3}[x+\frac{{\tilde e}}{F}]\big) 
\end{equation}where $\alpha$ and $\beta$ are two real constants to be determined. The wave functions should vanish at the frontiers of the potential well, $\varphi(0)=\varphi(L_x)=0$. These conditions help on the one hand to fix the values of $\alpha$ and $\beta$, and on the other hand to write down the secular equation that gives the set of discrete eigenvalues, ${\tilde e}$, 
\begin{equation} 
\label{secular1} 
\mathbf{Ai}\big(-{\tilde e}F^{-2/3}\big)\mathbf{Bi}\big(-L_x F^{1/3}-{\tilde e}F^{-2/3}\big)-\mathbf{Ai}\big(-L_x F^{1/3}-{\tilde e}F^{-2/3}\big)\mathbf{Bi}\big(-{\tilde e}F^{-2/3}\big)=0. 
\end{equation}We can equivalently express the Airy functions in terms of Bessel functions of fractional order \cite{abramovitch}, $\mathbf{J}_{1/3}$ and $\mathbf{J}_{-1/3}$, 
\begin{equation} 
\begin{array}{l} 
\mathbf{Ai}(-z)=\frac{\sqrt{z}}{3}\big(\mathbf{J}_{1/3}(\zeta)+\mathbf{J}_{-1/3}(\zeta)\big)\\ 
\mathbf{Bi}(-z)=\sqrt{\frac{z}{3}}\big(\mathbf{J}_{-1/3}(\zeta)-\mathbf{J}_{1/3}(\zeta)\big) 
\end{array} 
\end{equation}with $\zeta=\frac{2}{3}z^{3/2}$. The eigenvalue equation can then be rewritten as 
\begin{equation} 
\label{secular2} 
\mathbf{J}_{1/3}\big(\frac{2}{3}\frac{{\tilde e}^{3/2}}{F}\big)\mathbf{J}_{-1/3}\big(\frac{2}{3}\big[L_x F^{1/3}+{\tilde e}F^{-2/3}\big]^{3/2}\big)-\mathbf{J}_{-1/3}\big(\frac{2}{3}\frac{{\tilde e}^{3/2}}{F}\big)\mathbf{J}_{1/3}\big(\frac{2}{3}\big[L_x F^{1/3}+{\tilde e}F^{-2/3}\big]^{3/2}\big)=0. 
\end{equation}The equations (\ref{secular1}) and (\ref{secular2}) can be solved numerically, but additional approximations may be applied for our purpose. Indeed, especially at weak electric field, most of the eigenvalues are such that ${\tilde e}^{3/2}\gg FL_x$. In this case, we can make use of an asymptotic expansion of the Bessel function \cite{abramovitch} (when $z\rightarrow +\infty$, $\mathbf{J}_{\mu}\sim \sqrt{2/\pi z}\cos(z-\mu \pi/2-\pi/4)$) that greatly simplifies Eq.~(\ref{secular2}) to 
yield 
\begin{equation} 
\label{secular3} 
\big[L_x F^{1/3}+{\tilde e}F^{-2/3}\big]^{3/2}-\big[{\tilde e}F^{-2/3}\big]^{3/2}=n\frac{3\pi}{2} 
\end{equation}with $n \in \mathbf{N}^*$. The latter equation is easier to solve than Eq.~(\ref{secular2}) and is very reliable for large $n$. In fact, for all the cases we have considered, the results obtained from Eq.~(\ref{secular3}) are in excellent agreement with the exact ones, even for small $n$. To give an example, we get for $FL_x=10^{-2}$ a relative error of $10^{-4}$ for the first eigenenergy, $n=1$. The error rapidly decreases with increasing values of $n$. Starting from Eq.~(\ref{secular3}), we may further assume that ${\tilde e}\gg FL_x$. We can then expand the first term of Eq.~(\ref{secular3}) in powers of $\frac{FL_x}{{\tilde e}}$. At first order one gets, as expected, the free electron spectrum ${\tilde e}\simeq (\frac{n\pi}{L_x})^2$. Details about the opposite limit where ${\tilde e}\ll FL_x$ can be found in Ref.~~\onlinecite{rabinovitch}. In any case, the spectrum of the system with electric field is given by 
 
\begin{equation} 
\epsilon_{n_y}({\cal E})=\frac{\hbar^2}{2 m}\big[ ({\tilde 
e}+\frac{\pi n_y}{L_y})^2\big]. 
\end{equation}Examples are shown in Fig.~2, for $L_x=10L_y$. The two lowest bands, $n_y=1$ and $n_y=2$, are shown  with and without electric field. One can notice that the distortion of the bands induced by the electric field, depends on the energy or quantum number $n_x$: it is rather strong at low energies (small $n_x$) where a change of curvature can be seen in our example, and turns into a uniform shift at higher energies (large $n_x$). 
 
At zero temperature, the ground state of the system with $N_e$ 
electrons is determined by filling up successively the 
one-electron eigenstates according to the Pauli principle. In analogy with classifications done in atomic physics, it may 
be characterized by a set of occupation numbers, $A_{n_y}(E)$, that give 
the number of electrons in band $n_y$ at a particular electric 
field. The ground state may then be characterized by the product 
$\prod_{n_y,A_{n_y}\ne 0}n_y^{A_{n_y}}$ meaning that the band $n_y$ is 
populated by $A_{n_y}$ electrons; it gives a representation of 
the ground-state electronic configuration. Since the magnitude 
of the perturbation is not the same for every state $\varphi(x)$, 
one may expect the electric field to be able to induce changes in 
the set of occupation numbers. Such an example is seen in Fig.~2, 
where we have considered 34 electrons, 17 with spin up and 17 
with spin down. Without electric field, only states in the lowest 
band are doubly occupied (this is represented by the open 
diamonds in Fig~2) and in such a way that the lowest unoccupied 
level is in the second band. With an electric field,  the 
perturbation of the lowest level of the second band is larger 
than the perturbation of the highest occupied level of the first 
band. Consequently, an abrupt change in the set of occupation 
number occurs once the electric field becomes larger than a 
critical field, $E_c$, for which the lowest occupied and the 
highest unoccupied levels become degenerate. The value of this 
critical field dramatically depends on the geometry of the box: 
we obtain for instance, $E_c\simeq0.1\frac{\hbar^2}{2m|q|}$ for 
$L_x=10L_y$ and $E_c\simeq0.01\frac{\hbar^2}{2m|q|}$ for 
$L_x=9.8L_y$. For the case shown in Fig.~2, the ground state is 
characterized by the product $1^{34}$, for $E< E_c$, and 
$1^{32}2^2$, for $E \ge E_c$. 
 
Since a particular band, characterized by $n_y$, is associated 
with a particular function 
$\xi_{n_y}(y)=\sqrt{\frac{2}{L_y}}\sin(\frac{\pi n_y}{L_y}y)$, a 
change in the occupation numbers, $A_{n_y}$, corresponds a change 
in the electron density in the direction perpendicular to the 
applied electric field. This is well seen from the transverse 
electron density, $\rho ^T(y,E)$, which is the density with the 
$x$ variable integrated out

\begin{equation} 
\rho ^T(y,E)=\int^{L_x}_0dx \rho(x,y,E)=\frac{4}{N_eL_y}\sum_{n_y}A_{n_y}(E)\sin^2{\frac{\pi n_y}{L_y}y} 
\end{equation}where the upperscript $T$ is for {\it Transverse}. This quantity is expressed by the sum of the occupation numbers, $A_{n_y}$, weighted by the square of the corresponding transverse wavefunction, $\xi_{n_y}$. An example is shown in Fig.~3, where we have plotted the change of transverse density defined as the difference of $\rho ^T(y,E)$ with and without electric field 
 
\begin{equation} 
\delta\rho^T(y,E)=\rho ^T(y,E)-\rho 
^T(y,0)=\frac{4}{N_eL_y}\sum_{n_y}\left[A_{n_y}(E)-A_{n_y}(0)\right]\sin^2{\frac{\pi 
n_y}{L_y}y}. 
\end{equation} Since $A_{n_y}(E)=A_{n_y}(0)$ for $E< E_c$, this is a discontinuous function of the electric field, non-vanishing for $E>E_c$ only. In our example ($L_x=10L_y$, $N_e=34$), at $E=E_c=\frac{0.1}{|q|}\frac{\hbar^2}{2m}$ two electrons jump from the band $n_y=1$ to the band $n_y=2$ inducing the changes seen in Fig.~3: electrons leave the center of the dot to reach the two edges parallel to the applied electric field. Of course, this jump produces also a decrease of the longitudinal electric-polarization i.e. in the $x$-direction, an example is given in the next section. 
 
\section{Anharmonic oscillators with an applied electric field} 
As a next example, we consider electrons within a two dimensional 
box as in the previous section, but with a more realistic harmonic confinement. 
We assume the potential to be asymmetric (rectangular box) 
and we add a weak anharmonic term in the direction of the applied 
electric field. The corresponding one-particle time-independent 
Schr\"odinger equation is 
\begin{equation} 
\label{anharmonic-oscillator} 
-\frac{\hbar^2}{2m}\nabla^2\Psi(x,y)+[\frac{1}{2}m\omega_x^2 x^2+\frac{1}{2}m\omega_y^2 y^2+\lambda x^4+q E x]\Psi(x,y)=\epsilon \Psi(x,y) 
\end{equation}where we assume $\omega_x<\omega_y$ and $\lambda>0$. In practice, due to irregularities in the confinement potential anharmonicity 
always exists. Another alternative to justify this term in the model (\ref{anharmonic-oscillator}), would be to consider it to have a more fundamental origin, resulting from a renormalization procedure of a more realistic model including Coulomb interaction, for instance. In any case, the anharmonic term introduces a 
field-dependent perturbation, function of the unperturbed energies, which does not exist in the pure harmonic case and which is crucial for our purpose, as will be explained in the following two subsections. 
 
\subsection{Harmonic oscillators in an electric field} 
 
First, we neglect the anharmonic term and consider the case 
without electric field, $\lambda=E=0$. In this case one recovers the very well 
known two dimensional harmonic oscillator problem. The 
wave-functions are written as a product, 
$\Psi_{n_x,n_y}(x,y)=\varphi_{n_x}(x)\varphi_{n_y}(y)$, with 
 
\begin{equation} 
\label{wavefunctionho} 
\varphi_{n_u}(u)=(\frac{\beta_u^2}{\pi})^{1/4}\frac{1}{\sqrt{2^{n_u}n_u!}}e^{-\beta_u^2u^2/2}H_{n_u}(\beta_uu) 
\end{equation}where $u=x/y$, $\beta_u=\sqrt{\frac{m\omega_u}{\hbar}}$, the inverse of the effective Bohr radius, and $H_{n_u}$ is the Hermite's polynomial of order $n_u$. The corresponding eigenenergies  are 
 
\begin{equation} 
\epsilon_{n_x,n_y}=(n_x+\frac{1}{2})\hbar \omega_x+(n_y+\frac{1}{2})\hbar \omega_y 
\end{equation}where $n_x$ and $n_y$ are two integers \cite{cohen}. 
 
With an applied electric field but without anharmonicity, 
$\lambda=0$ and $E \ne 0$, the overall spectrum of the harmonic 
oscillator is shifted by $-\frac{q^2E^2}{2m\omega_x^2}$ which 
corresponds to a shift  of the center of mass by $-\frac{q 
E}{m\omega_x^2}$, in the $x$-direction. The eigenfunctions become \cite{cohen} 
 
\begin{equation} 
\label{wavefunctionhoE} 
\varphi_{n_x}(x,E)=(\frac{\beta_x^2}{\pi})^{1/4}\frac{1}{\sqrt{2^{n_x}n_x!}}e^{-\beta_x^2(x+\frac{qE}{m\omega_x^2})^2/2}H_{n_x}(\beta_x(x+\frac{qE}{m\omega_x^2})). 
\end{equation}The harmonic confinement and the shift induced by an homogeneous electric field are schematically shown in Fig.~4. Since the perturbation induced by the field is the same for all eigenenergies, there is no change in the set of occupation numbers, $A_{n_y}$, and thereby, in the electron density. This property is very peculiar to the harmonic potential confinement. The situation is very different with anharmonicity as it is shown in the next 
subsection. 
 
\subsection{Effects of anharmonicity} 
 
We assume $\lambda$ very small compared to the other energy 
scales, $\lambda/\beta_x^4 \ll \hbar \omega_u$ ($u=x/y$). It has been known that 
the perturbative expansion in $\lambda$ is not convergent but 
asymptotic \cite{hioe}. Yet, the perturbative results may 
approximate the exact one if they are restricted to low orders. 
Here we assume $\lambda$ sufficiently small to consider the 
anharmonic term at first order in perturbation theory only. Although the range of validity of this treatment is very restricted, our purpose is mainly qualitative and aims at pointing out basic 
mechanisms rather than reaching a quantitative description of the phenomenon that would require more sophisticated approaches \cite{hioe}. We believe that the effects describe here occur at any $\lambda$. We calculate the correction 
\begin{equation} 
\label{perturbation} \Delta_{n_x}(E)=\lambda 
\int^{+\infty}_{-\infty}dx \varphi_{n_x}^2(x,E)x^4=\lambda 
\int^{+\infty}_{-\infty}dx 
\varphi_{n_x}^2(x,0)(x-\frac{qE}{m\omega_x^2})^4. 
\end{equation} Exploiting the well known recurrence relation for Hermite 
polynomials, $xH_n(x)=nH_{n-1}(x)+\frac{1}{2}H_{n+1}(x)$, and 
noting that the integrals with odd power of $x$ are null, we get 
 
\begin{equation} 
\label{perturbation2} 
 \Delta_{n_x}(E)=\frac{3}{2}\frac{\lambda \hbar^2}{m^2 
\omega_x^2}\big(n_x^2+n_x+\frac{1}{2}\big)+6\lambda\frac{\hbar 
q^2E^2}{m^3\omega_x^5}(n_x+\frac{1}{2}) 
\end{equation}plus an overall shift of the spectrum $\lambda\frac{q^4E^4}{m^4\omega_x^8}$ that plays no role here; thus we neglect it in the following. The last term of~(\ref{perturbation2}) is driven by anharmonicity and introduces an explicit dependence on the applied electric field, $E$, and on the quantum number, $n_x$: an applied electric field is able to change the electronic configurations of the ground state if anharmonicity is included, like in the case studied in section II. We illustrate this property below by studying one particular example. 
 
Using Eq.~(\ref{perturbation2}), the energies of the different levels can be determined at 
first order in perturbation theory, omitting the two uniform shifts: 
$\tilde{\epsilon}_{n_x,n_y}=\epsilon_{n_x,n_y}+\Delta_{n_x}(E)$. 
We consider a system with four electrons, two spin up and two spin 
down, so that only the two occupied and the first unoccupied levels should be considered. We first introduce a parameter, $\delta$, which control the 
asymmetry of the harmonic confinement: $\omega_x=\omega-\delta$ 
and $\omega_y=\omega+\delta$. The energies of the three lowest 
levels including the anharmonic corrections (\ref{perturbation2}) 
are found to be 
\begin{equation} 
\label{lowest3levels} 
\left\{ 
\begin{array}{l} 
\tilde{\epsilon}_{0,0}(E)=\hbar \omega+\frac{3}{4}\frac{\lambda}{\beta_x^4}+3\lambda\frac{x_E^2}{\beta_x^2}\\ 
\tilde{\epsilon}_{1,0}(E)=\hbar( 2\omega-\delta)+\frac{15}{4}\frac{\lambda}{\beta_x^4}+9\lambda\frac{x_E^2}{\beta_x^2}\\ 
\tilde{\epsilon}_{0,1}(E)=\hbar( 2\omega+\delta)+\frac{3}{4}\frac{\lambda}{\beta_x^4}+3\lambda\frac{x_E^2}{\beta_x^2}\\ 
\end{array} 
\right . 
\end{equation}where we have introduced $x_E=-\frac{qE}{m\omega_x^2}$, the shift of the center of mass induced by the electric field \cite{cohen}. At a critical field, $E_c$, there is a change of the ground state electronic configuration, $\prod_{n_y} n_y^{A_{n_y}}$, from $0^4$ to $0^21^2$. The critical field is such that $\tilde{\epsilon}_{1,0}(E_c)=\tilde{\epsilon}_{0,1}(E_c)$;  within our approximation it is given by 
\begin{equation} 
\label{criticalfield} 
E_c=\frac{\hbar \omega_x \beta_x}{|q|}\sqrt{\frac{2\hbar\delta\beta_x^4-3\lambda}{6\lambda}}. 
\end{equation}This change abruptly modifies the electron density in both direction, $x$ and $y$, which can be calculated, at our level of approximation, 
by using the wave functions (\ref{wavefunctionho}) and 
(\ref{wavefunctionhoE}). 
 
As already discussed for the case of hard-wall boundary confinement, for $E\ge E_c$ the electronic cloud suddenly swells in the y-direction, perpendicularly to the applied field. The change of density in the transverse direction is given by 
 
\begin{equation} 
\delta\rho^T(y,E)=\frac{1}{2}\sqrt{\frac{\beta_y^2}{\pi}}e^{-\beta_y^2y^2}(2\beta_y^2y^2-1)\Theta(E-E_c) 
\end{equation}where $\Theta(x)$ is the usual step function 
\begin{equation} 
\Theta(x)=\left\{\begin{array}{l}0,\quad if\quad x<0\\ 
1,\quad if \quad x\ge0\end{array}\right .. 
\end{equation}Results for our example are shown in Fig.~5: electrons are transfered from the center to the edges of the rectangular dot, parallel to the direction of the field. 
 
The center of mass of the electronic system is continuously shifted by $x_E$ in the longitudinal direction (the x-direction) when the electric field is tuned (Fig.~4). In addition, at the transition, $E=E_c$, the internal structure of the electronic cloud is modified in this direction too. To characterize the changes, we may consider the difference between the longitudinal electron density i.e. the electron density where the y-coordinate has been integrated out, taken at two different values of the field, $E_c^+$ and $E_c^-$, slightly above and below the transition, $E_c$, respectively: 
 
\begin{equation} 
\delta\rho^L(x,E_c^+,E_c^-)=\rho^L(x,E_c^+)-\rho^L(x,E_c^-)=\frac{1}{2}\sqrt{\frac{\beta_x^2}{\pi}}e^{-\beta_x^2(x-x_E)^2}(1-2\beta_x^2(x-x_E)^2) 
\end{equation}where $\rho^L(x)=\int_{-\infty}^{+\infty}dy\rho(x,y)$, $\rho$ the electron density. Results for our example are shown in Fig.~6 where we have taken $\beta_xx_E=0.5$: electrons are transfered from the edges to the center of the dot. In conclusion, at the transition, the electronic cloud grows in the transverse direction ($y-$direction) but is slightly contracted in the longitudinal direction ($x-$direction). For Quantum Dots in GaAs, $\hbar\omega_x$ is about a few meV that corresponds to an effective Bohr radius, $1/\beta_x$, of a few tens of nanometers; with these conditions a typical value for the critical field, at which the transition should occur, can be estimated with Eq.~(\ref{criticalfield}), to be $E_c\approx 10^{-5} V.m^{-1}$, a value commonly used in most experimental situations. 
 
Before closing the section, one should say that in order to observe this jump of electron density, the number of electrons in the dot has to be adjusted in a way to reduce the gap between the highest occupied level and the lowest unoccupied level; for our particular example, the case with 4 particles is optimal in this respect.

\section{Two laterally coupled quantum dots} 
 
The effect pointed out above should have consequences in 
various experimental situations. In this last section we propose 
an experimental set-up that should enable to detect it. It is inspired by a proposal made in the late 90s to use 
semiconductor quantum dots as building blocks for hypothetical future 
quantum computers \cite{loss,burkard,dsasarma}. Following 
Ref.~~\onlinecite{burkard}, we consider two lateral quantum dots weakly coupled (see Fig.~6), described by the following model 
 
\begin{equation} 
\label{Htwodots} 
\hat{H}=\sum_{\alpha}\big(-\frac{\hbar^2}{2m}\nabla_{\alpha}^2+V(x_{\alpha},y_{\alpha})+\lambda 
x^4_{\alpha}+qEx_{\alpha} \big)+\frac{1}{2}\sum_{\alpha \ne 
\beta}\frac{q^2}{\kappa|\mathbf{r_{\alpha}}-\mathbf{r_{\beta}}|} 
\end{equation}with 
\begin{equation} 
V(x_{\alpha},y_{\alpha})=\frac{1}{2}m\omega_x^2x^2_{\alpha}+\frac{1}{2}m\omega_y^2\frac{(y^2_{\alpha}-a^2)^2}{4a^2} 
\end{equation}where the summation is over the particle indices, $\alpha$ and $\beta$. The first term in brackets in Eq.~(\ref{Htwodots}) describes free electrons in a confined potential; in particular the term $V(x,y)$ represents a typical double well, which is well approximated by an harmonic potential when the distance $2a$ between the dots is large compared to the effective Bohr radius $1/\beta_y$. Then, with respect to Ref.~~\onlinecite{burkard} we add here the anharmonic contribution and an applied electric field perpendicular to the line joining the centers of the two quantum dots. The second term of Eq.~(\ref{Htwodots}) is the Coulomb interaction supposed to be long-ranged; $\kappa$ is the dielectric constant and $\mathbf{r_{\alpha}}$ is the vector coordinates of the electron $\alpha$, in the 3D space. 
 
We assume that the two dots are weakly coupled, meaning that the distance 
between them, $2a$, is large ($2a> 1/\beta$); in this case, the 
tunneling probability of an electron from one dot to the other is small and the two dots can be considered, at first approximation, as being uncoupled. The analysis that we have done in the previous section for one single dot is then relevant for the coupled system. We consider the case with a small odd number of particles in each dot, leaving one unpaired electron in dot 1 and dot 2. In this case, under our assumptions, the low-energy properties of this system can be 
described by an effective Heisenberg Hamiltonian for spin $1/2$ \cite{loss,burkard,dsasarma}, 
\begin{equation} 
\label{heisenberg} 
\hat{H}_{Heis}=J\mathbf{S_1}.\mathbf{S_2} 
\end{equation}where the spin operators $\mathbf{S_1}$ and 
$\mathbf{S_2}$ are associated to the spin of the unpaired electrons 
localized in dot 1 and dot 2, respectively. The effective 
exchange integral, $J$, can be estimated starting from the more microscopic model (\ref{Htwodots}): it is defined as the energy difference 
between the lowest triplet and singlet states. Its magnitude 
depends on the Coulomb potential and the interdot tunneling 
probability. Our basic idea may be summarized as follows. We have seen, in the previous section, that an applied electric field along the x-direction may change abruptly the electron density in the y-direction: in each dot, at a critical field, $E_c$, electrons are transfered from the center to the boundaries. This phenomenon should contribute to increase the overlap between the two 
electronic clouds centered in dot 1 and dot 2, and therefore it should modify the magnitude of the effective exchange integral of Eq.~(\ref{heisenberg}).  Consequently, depending on the magnitude of the electric field one expects the system to be described by two 
different Heisenberg models, with different exchange integrals, 
$J^<$, for $E<E_c$, and $J^>$, for $E>E_c$. In other words, the exchange interaction of the effective spin model is a function of the electric field that can be written as (see Fig.~9) 
\begin{equation} 
J(E)=J^<\Theta(E_c-E)+J^>\Theta(E-E_c). 
\end{equation}In the following, we 
determine these exchange integrals using a generalization of the Heitler-London method for the hydrogen molecule, to a number of electrons larger than two \cite{mcweeny}; this type of calculation is valid for large interdot 
distances. Within this approximation we first consider the 
solutions for the isolated non-interacting dots (section III) before using them 
to build the proper combinations of Slater determinants to describe 
the whole system with interaction. 
 
The ground state of each isolated dot has been studied in the previous section, at first order in perturbation of $\lambda$. We start by introducing new fermionic operators, $L^{\dagger}_{n_x,n_y,\sigma}$ ($R^{\dagger}_{n_x,n_y,\sigma}$) 
which create one electron with spin $\sigma$ ($\uparrow$ or 
$\downarrow$) in the orbital state labeled by the quantum numbers $n_x$ 
and $n_y$ in the left (right) dot; they are associated with the wavefunctions $\varphi_{n_x}(\beta_x(x-x_E))\varphi_{n_y}(\beta_y(y+a))$ ($\varphi_{n_x}(\beta_x(x-x_E))\varphi_{n_y}(\beta_y(y-a))$ and energies $\tilde{\epsilon}_{n_x,n_y}$. We specify our study to the case with three electrons in each dot with the same number of spin up and down ($S_z=0$). In this case, it is sufficient to consider only the three lowest levels (\ref{lowest3levels}), and the gap between occupied and unoccupied states is reduced. Using the previous results, one concludes that the electronic configuration of the ground state changes from $0^3$, for $E<E_c$, to $0^21^1$, for $E>E_c$. This is depicted schematically in Fig~~8, where only one particular configuration is shown. Starting from these solutions for isolated dots, we build for each of these two cases, the singlet and triplet states of the whole system. For $E<E_c$, we get 
 
\begin{equation} 
\Psi_{\pm}^<=\frac{1}{\sqrt{2}}\big(\Psi_{\uparrow \downarrow}^< \pm \Psi_{\downarrow \uparrow}^< \big) 
\end{equation}with 
\begin{equation} 
\begin{array}{c} 
\Psi_{\uparrow \downarrow}^< = L^{\dagger}_{0,0,\downarrow}L^{\dagger}_{0,0,\uparrow}R^{\dagger}_{0,0,\downarrow}R^{\dagger}_{0,0,\uparrow}R^{\dagger}_{1,0,\downarrow}L^{\dagger}_{1,0,\uparrow}|0> \\ 
\Psi_{\downarrow \uparrow}^< = L^{\dagger}_{0,0,\downarrow}L^{\dagger}_{0,0,\uparrow}R^{\dagger}_{0,0,\downarrow}R^{\dagger}_{0,0,\uparrow}L^{\dagger}_{1,0,\downarrow}R^{\dagger}_{1,0,\uparrow}|0> 
\end{array} 
\end{equation}and in the same way, we get for $E>E_c$ 
 
\begin{equation} 
\Psi_{\pm}^>=\frac{1}{\sqrt{2}}\big(\Psi_{\uparrow \downarrow}^> \pm \Psi_{\downarrow \uparrow}^> \big) 
\end{equation}with 
\begin{equation} 
\begin{array}{c} 
\Psi_{\uparrow \downarrow}^> = L^{\dagger}_{0,0,\downarrow}L^{\dagger}_{0,0,\uparrow}R^{\dagger}_{0,0,\downarrow}R^{\dagger}_{0,0,\uparrow}R^{\dagger}_{0,1,\downarrow}L^{\dagger}_{0,1,\uparrow}|0> \\ 
\Psi_{\downarrow \uparrow}^> = L^{\dagger}_{0,0,\downarrow}L^{\dagger}_{0,0,\uparrow}R^{\dagger}_{0,0,\downarrow}R^{\dagger}_{0,0,\uparrow}L^{\dagger}_{0,1,\downarrow}R^{\dagger}_{0,1,\uparrow}|0> 
\end{array} 
\end{equation}where the $+$ sign is for the singlet, the $-$ being for the triplet, and $|0>$ is the vacuum. 
 
Next, after evaluating the triplet and singlet energy for the two different cases, 
 
\begin{equation} 
\label{energies} 
\epsilon_{\pm}^{<,>}=\frac{<\Psi_{\pm}^{<,>}|\hat{H}|\Psi_{\pm}^{<,>}>}{<\Psi_{\pm}^{<,>}|\Psi_{\pm}^{<,>}>} 
\end{equation}we can determine the two different exchange integrals, $J^<$ and $J^>$ of the effective Heisenberg model (\ref{heisenberg}) 
 
\begin{equation} 
\label{exchange} 
\begin{array}{c} 
J^<=\epsilon^<_- - \epsilon^<_+ \\ 
 
J^>=\epsilon^>_- - \epsilon^>_+ 
\end{array} 
\end{equation}given by the difference between the triplet and the singlet energy \cite{burkard}.

The wavefunctions used to describe the systems are not all orthogonal since 
 
\begin{equation} 
\label{overlap} 
S_{n_y,n'_y}=\int^{+\infty}_{-\infty}\varphi_{n_y}(y+a)\varphi_{n'_y}(y-a) \ne 0. 
\end{equation}This property makes the calculation of the energy~(\ref{energies}) more cumbersome. However, using the wavefunction~(\ref{wavefunctionho}), we can estimate that the overlap behaves as $S_{n_y,n'_y}=($Polynomial of $a)\times e^{-\beta_y^2a^2}$. Since we assume $a$ to be relatively large with respect to $1/\beta_y$, a reasonably good approximation consists to keep in the calculation, terms of the type $S_{n_y,n'_y}^m$ with $m\le2$. With this simplification, after lengthy but straightforward calculations, we obtain
 
\begin{equation} 
\label{exchangeD} 
J^<=e^{-2\beta_y^2a^2}\big[\frac{q^2}{\kappa}R_K^< 
\mathbf{K}\big(\frac{\sqrt{\beta_y^2-\beta_x^2}}{\beta_y}\big)+\frac{q^2}{\kappa}R_E^< 
\mathbf{E}\big(\frac{\sqrt{\beta_y^2-\beta_x^2}}{\beta_y}\big)+\frac{3}{2}\hbar\omega_y\big(1-3\beta_y^2a^2\big)\big] 
\end{equation}with 
\begin{equation} 
\left\{ 
\begin{array}{l} 
R^<_K=\sqrt{\frac{2}{\pi}}\beta_x\frac{12\beta_x^4+\beta_y^4-15\beta_x^2\beta_y^2}{2(\beta_y^2-\beta_x^2)^2}\\ 
\\ 
R^<_E=\sqrt{\frac{2}{\pi}}\beta_x\frac{6\beta_y^4-5\beta_x^2\beta_y^2}{(\beta_y^2-\beta_x^2)^2} 
\end{array} 
\right . 
\end{equation}and 
\begin{equation} 
\label{exchangeU} 
J^>=e^{-2\beta_y^2a^2}\big[\frac{q^2}{\kappa}R_K^> 
\mathbf{K}\big(\frac{\sqrt{\beta_y^2-\beta_x^2}}{\beta_y}\big)+\frac{q^2}{\kappa}R_E^> 
\mathbf{E}\big(\frac{\sqrt{\beta_y^2-\beta_x^2}}{\beta_y}\big)+\hbar 
\omega_y(2\beta_y^2a^2-1)(-\frac{3}{2}+\frac{7}{2}\beta_y^2a^2-9\beta_y^4a^4)\big] 
\end{equation}with 
\begin{equation} 
\left \{ 
\begin{array}{l} 
R^>_K=\sqrt{\frac{2}{\pi}}\beta_x\frac{-\beta_x^4+(48\beta_y^4a^4-48\beta_y^2a^2+15)\beta_y^4+(8\beta_y^2a^2+2)\beta_x^2\beta_y^2}{2\beta_y^2(\beta_y^2-\beta_x^2)}\\ 
\\ 
R^>_E=\sqrt{\frac{2}{\pi}}\beta_x\frac{\beta_x^2+(24\beta_y^4a^4-20\beta_y^2a^2+5)\beta_y^2}{\beta_y^2-\beta_x^2} 
\end{array} 
\right . 
\end{equation}where $\mathbf{K}$ and $\mathbf{E}$ are the complete elliptic integrals of the first and second kind, respectively, defined as \cite{abramovitch} 
\begin{equation} 
\mathbf{K}(k)=\int^{\pi/2}_0 \frac{d\theta}{\sqrt{1-k^2\sin^2\theta}}, \quad \mathbf{E}(k)=\int^{\pi/2}_0d\theta\sqrt{1-k^2\sin^2\theta}. 
\end{equation}

Although, the two expressions (\ref{exchangeD}) and (\ref{exchangeU}) are only valid in the regime of weak anharmonicity, $\lambda/\beta_x^4<<\hbar\omega_{x/y}$, and large interdot distances, $\beta_ya>1$,  they allow us to illustrate our purpose. Indeed, they show that the exchange integral can be significantly modified by applying an electric field  perpendicular to the axis joining the two dots. Taking relevant numerical values for dots in GaAs \cite{burkard}, $\hbar\omega_y=3$meV, which corresponds to $1/\beta_y=20$nm, and $\beta_yq^2/\kappa/\hbar\omega_y=2.1$ and specifying our system with $\beta_ya=1.3$, we get for $\beta_x/\beta_y=0.9$, $J^<=0.15$meV and $J^>=0.6$meV and, for $\beta_x/\beta_y=0.95$, $J^<=0.19$meV and $J^>=3.13$meV. We estimate the critical field in the latter case by using Eq.~(\ref{criticalfield}): assuming $\lambda/\beta_x^4=0.18$meV, we obtain $E_c=35.10^{-7}$Vm$^{-1}$. In both cases, the energy difference between the singlet and triplet states is large enough to be probed experimentally, as it was done in Ref.~~\onlinecite{lee} for two coupled dots with an exchange energy tuned by an applied magnetic field as suggested in Ref.~~\onlinecite{burkard}. 
 
D. Loss and D. P. Di Vincenzo proposed to use a system of two weakly coupled quantum dots as possible realization of quantum gate \cite{loss}. With one unpaired electron in each dot, the system is described by an effective Heisenberg model such as (\ref{heisenberg}). By changing the value of the exchange energy with external sources in such a way that $\int^{\tau}_0dtJ(t)/\hbar=\pi$, it would be possible to perform a {\it swap} operation that exchanges the quantum state of dot 1 and dot 2. When combined with single-qubit operations, this operation can be used to build a quantum XOR gate that was earlier recognized to be a universal quantum gate \cite{divincenzo}: a XOR along with single-qubit operations may be assembled to do any quantum computation. Thus, the study of quantum gates is reduced to the study of the exchange integral of the two dots system and in the way it can be controlled experimentally. Two main approaches have been proposed up to now to control $J(t)$: (i) by a local magnetic field perpendicular to the plane of the dots and, (ii) by a local electric field parallel to the axis joining the center of the two dots. Our proposal offers another possibility. With respect to the method described in Ref.~~\onlinecite{burkard}, the main advantages of the mechanism proposed here are (i) to let the average charges unchanged within the two dots, quantity that would be modified by an electric field applied along the $y$-axis (see Fig.~7); (ii) it is easier to apply a local electric field than a local magnetic field.

As a remark, we would like to point out that there is no need to consider identical dots. Both the number of particles within each dot and the geometry of the dots might be different. In such situation, the exchange integral, $J(E)$, should show two discontinuities instead of one at two different critical field, $E_{c1}$ and $E_{c2}$, characterizing the electronic changes in the dot $1$ and dot $2$, respectively. 
 
\section{conclusion} 
 
To conclude, we summarize our main results. We have considered systems of free-electrons trapped in a small region of space. Two confining potentials were studied as examples: hard wall boundary conditions and harmonic potential with an additional perturbative $x^4$ term. In both cases, an electric field was shown to cause abrupt changes in the electron density at a critical value, $E_c$. These changes occur not only in the direction parallel to the field but also in the direction perpendicular to it; they can be related to finite size effects. Exploiting this outcome, we have then considered a system of two weakly coupled Quantum Dots and shown that an applied homogeneous electric field is able to change abruptly the difference in energy between the lowest singlet and triplet states. This property may be important in the context of quantum computation, since it could be exploited to realize a quantum gate. Our results were obtained by using deliberately oversimplified models, our aim being to point out basic mechanisms that should survive more realistic descriptions. Of course, to be more quantitative, more detailed models should be studied in the future. 
 
As a last remark, we have seen that the electric field induces degeneracy in the electronic spectrum at $E=E_c$. In this case, the Coulomb interaction should play a central role that could significantly modify our conclusions for fields close to the critical value: in particular, the total spin of the system may be changed. This kind of transitions was already studied, for instance, in Ref.~~\onlinecite{pleutin2}, for cylinders pierced by an Aharonov-Bohm flux, and in Ref.~~\onlinecite{oreg}, for single-wall carbon nanotubes with an inhomogeneous electric field.

\newpage

\begin{figure} 
\includegraphics[width=16cm,angle=0]{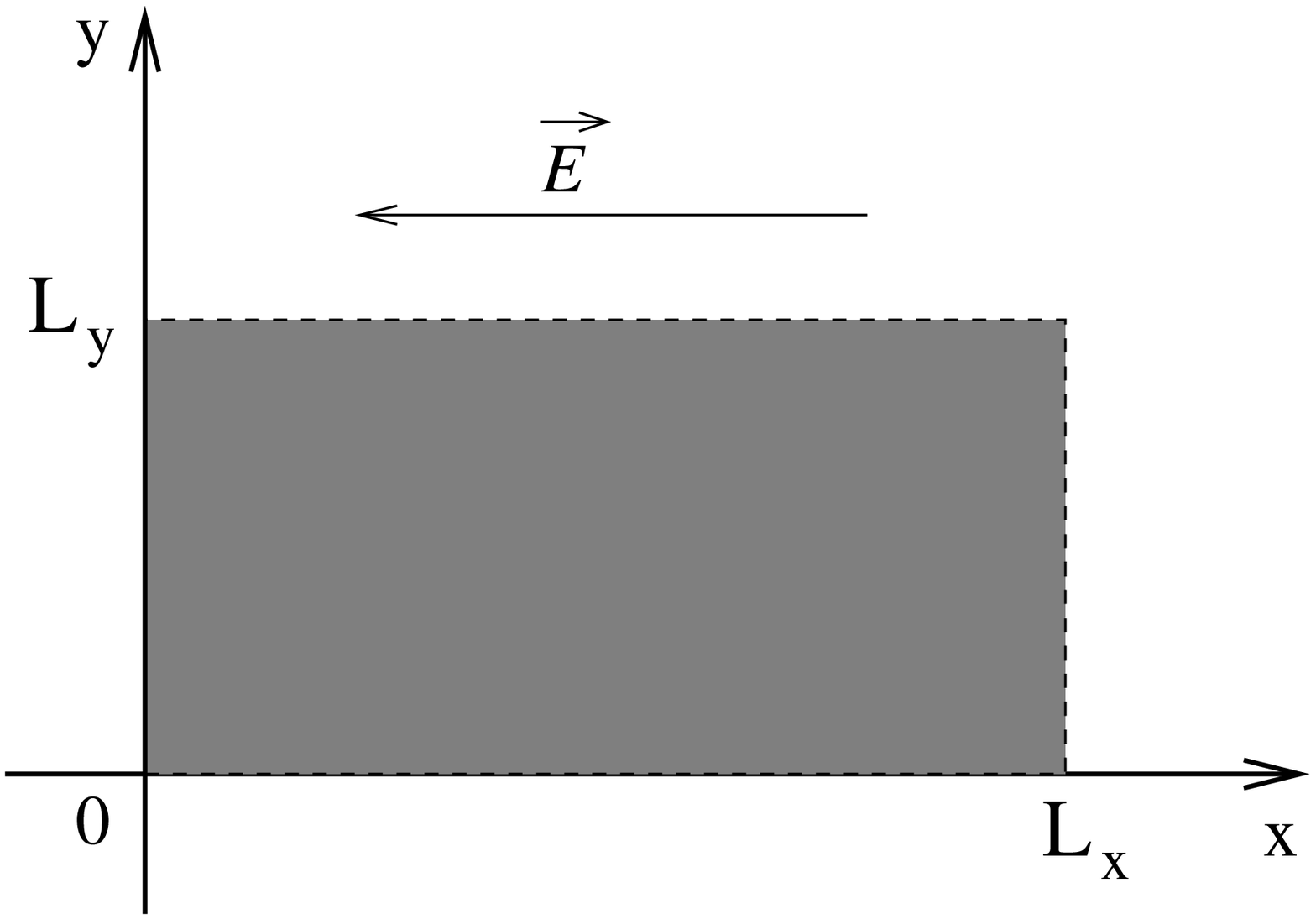} 
\caption{Rectangular potential well with infinite walls 
($L_x>L_y$). A uniform electric field, $E$, is applied along the 
long axis.} 
\end{figure} 
 
\begin{figure} 
\includegraphics[width=12cm,angle=-90]{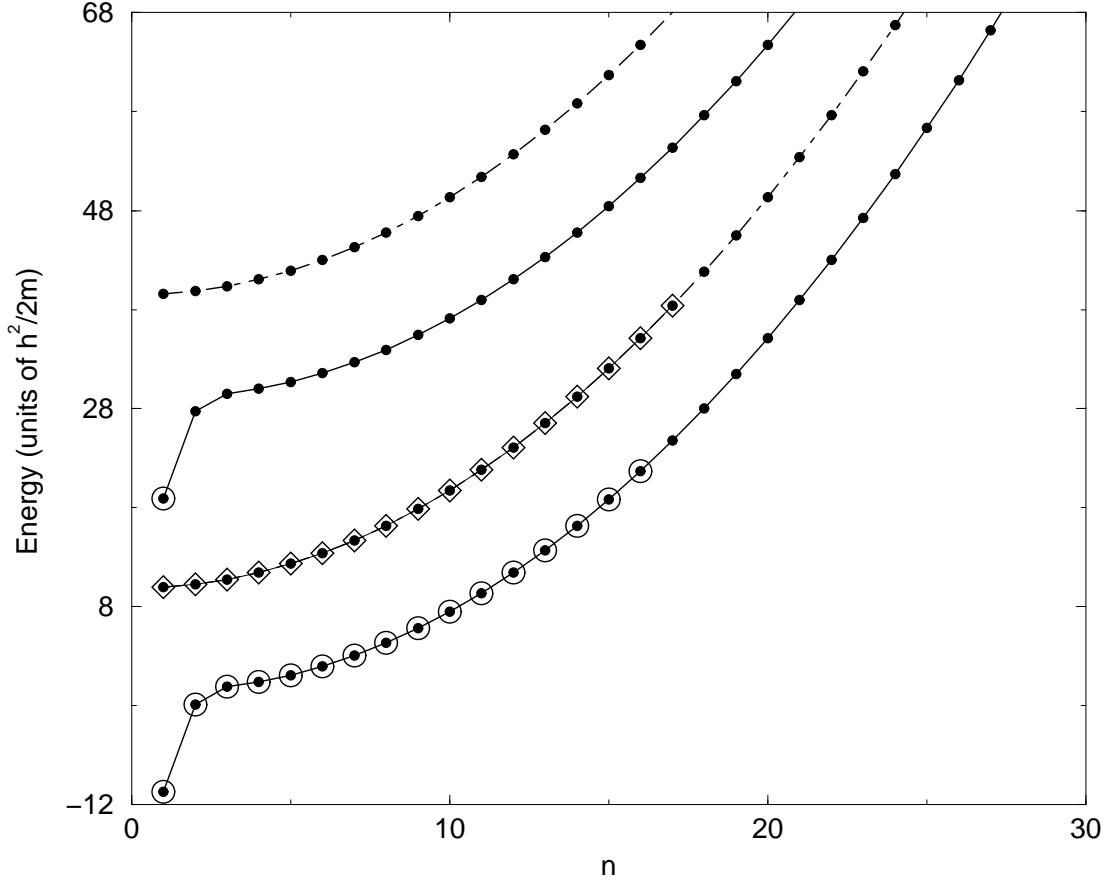} 
\caption{The two lowest bands, $n_y=1$ and $n_y=2$, of a system of free-electrons confined in a rectangular potential well such that $L_x=10L_y$. The long dashed curves are for $E=0$, the full curves for $E=\frac{0.1}{|q|}\frac{\hbar^2}{2m}$. The positions of the bands with electric field are shifted down by $\pi^2$, for clarity. In both cases, the dots denote the position of the discrete set of quantum states available due to finite sizes. They are labeled by the integer $n$, equivalent to $n_x$ in the case without electric field. The open diamonds and the open circles are the doubly occupied levels in the case with 34 electrons and $S_z=0$, without and with electric field, respectively. The electric field induces a change in the set of occupation numbers, the ground state changing from $1^{34}$ to $1^{32}2^2$ (see text): two electrons jump from the band $n_y=1$ to the band $n_y=2$.} 
\end{figure} 
 
\begin{figure} 
\includegraphics[width=12cm,angle=-90]{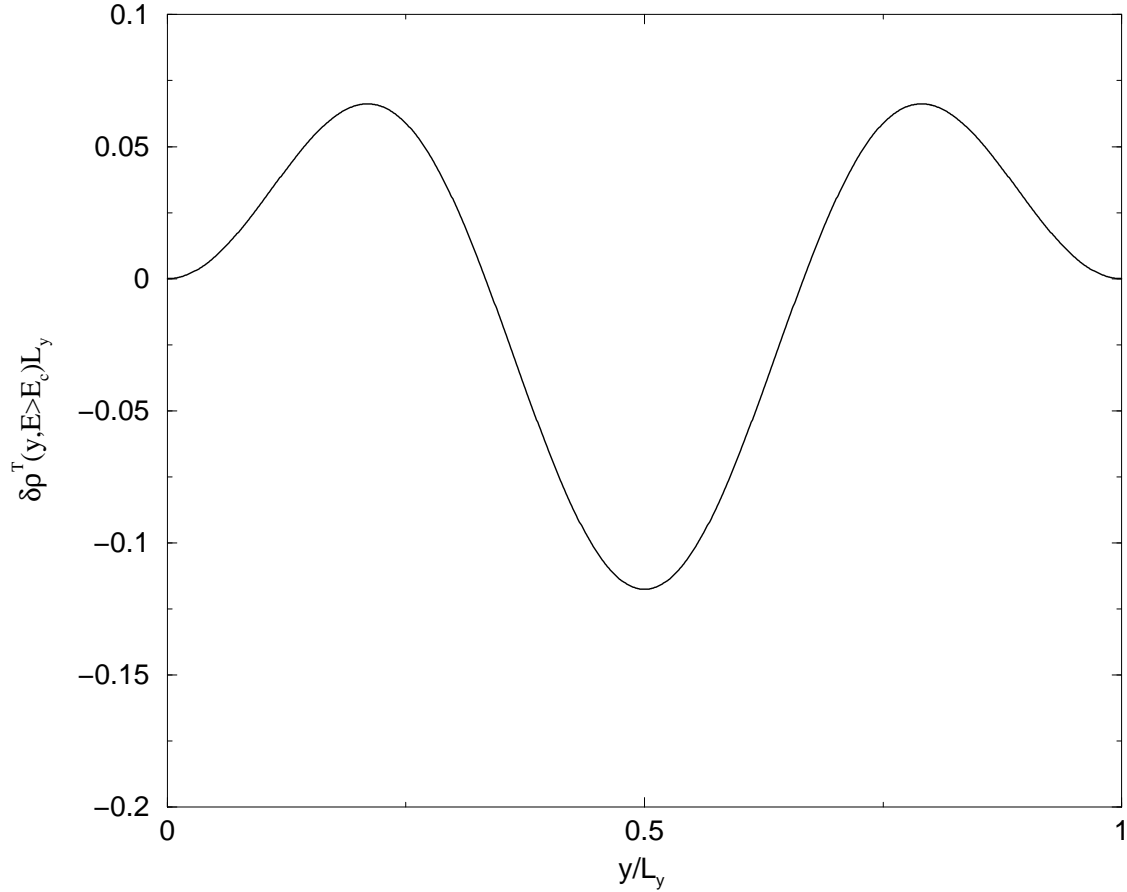} 
\caption{Changes in the transverse electron density, $\delta\rho^T(y,E>E_c)$, of a system of free-electrons confined in a rectangular box with hard-wall boundary conditions. At a critical value, $E_c$, an electric field induces abrupt changes in the set of occupation numbers, $A_{n_y}$, defining the ground state (see text) and, consequently, in the electron density in the direction perpendicular to the applied field. In this example, the ground state changes from $1^{34}$ to $1^{32}2^2$. The value of the critical field depends on the geometry of the well: it is $E_c\simeq0.1\frac{\hbar^2}{2m|q|}$ for $L_x=10L_y$ and $E_c\simeq0.01\frac{\hbar^2}{2m|q|}$ for $L_x=9.8L_y$, for instance.} 
\end{figure} 
 
\begin{figure} 
\includegraphics[width=14cm,angle=0]{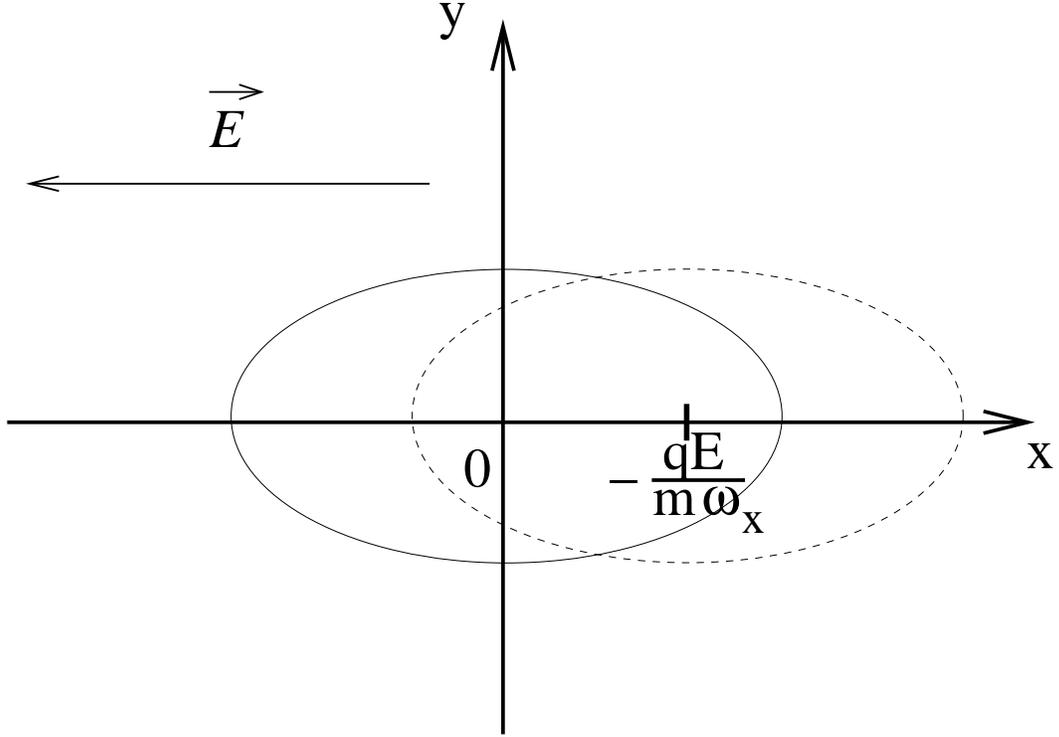} 
\caption{Rectangular quantum dot with harmonic confinement 
potential, such that $\omega_x<\omega_y$ (full line). An applied 
electric field along the $x$-axis shifts the position of the 
center of mass of the electronic system by $x_E=-\frac{qE}{m\omega_x}$ (dashed line).} 
\end{figure} 
 
\begin{figure} 
\includegraphics[width=14cm,angle=-90]{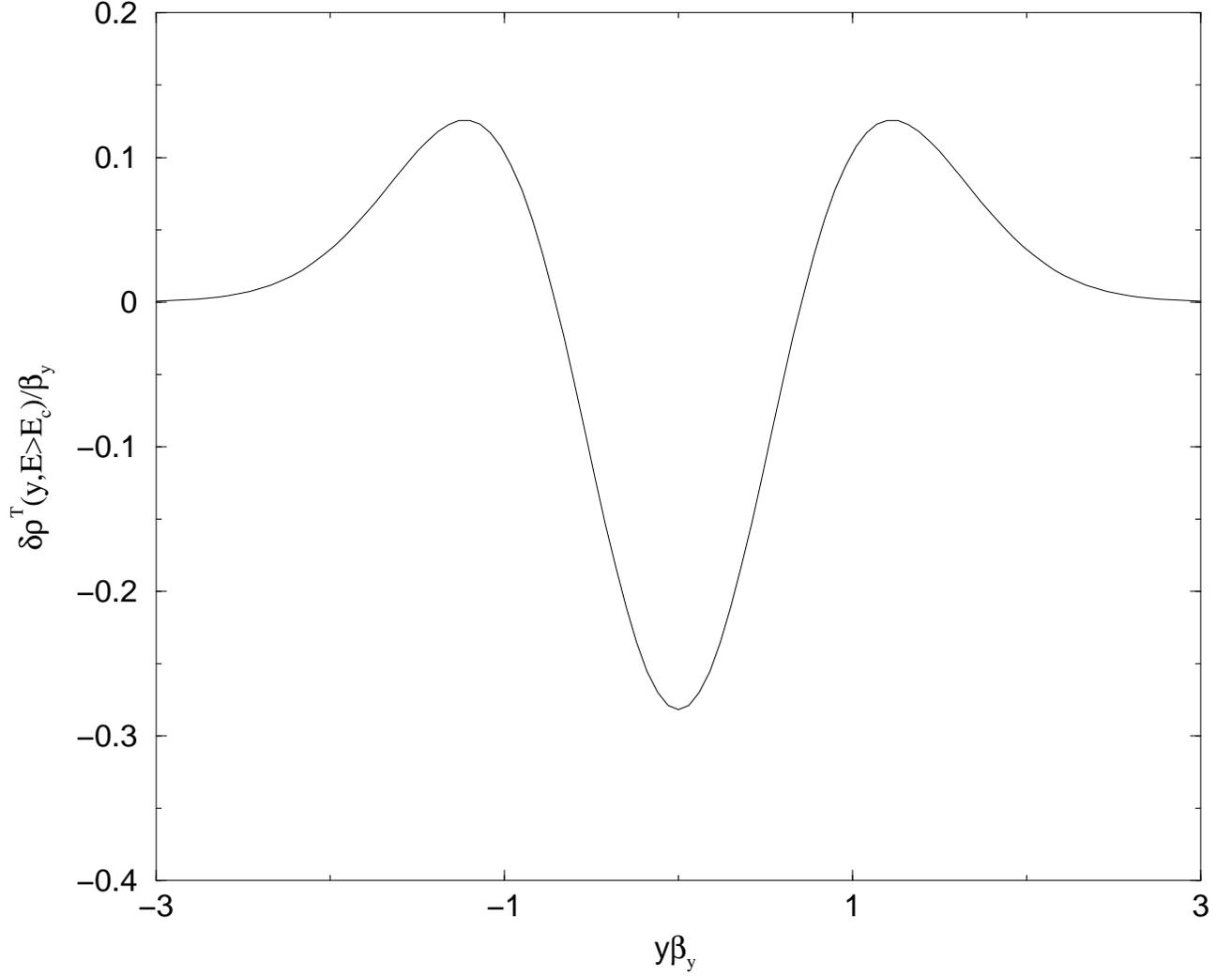} 
\caption{Changes in the transverse electron density, $\delta\rho^T(y,E>E_c)$, of free-electrons trapped in a harmonic potential with an additional weak anharmonic term, $\lambda x^4$. In our example, the ground state 
configuration, $\prod n_y^{A_{n_y}}$, changes from $0^4$ to 
$0^21^2$, at $E=E_c$. As a consequence of that, some electrons leave the 
center of the dot towards the edges parallel to the direction of 
the applied field.} 
\end{figure} 
 
\begin{figure} 
\includegraphics[width=14cm,angle=-90]{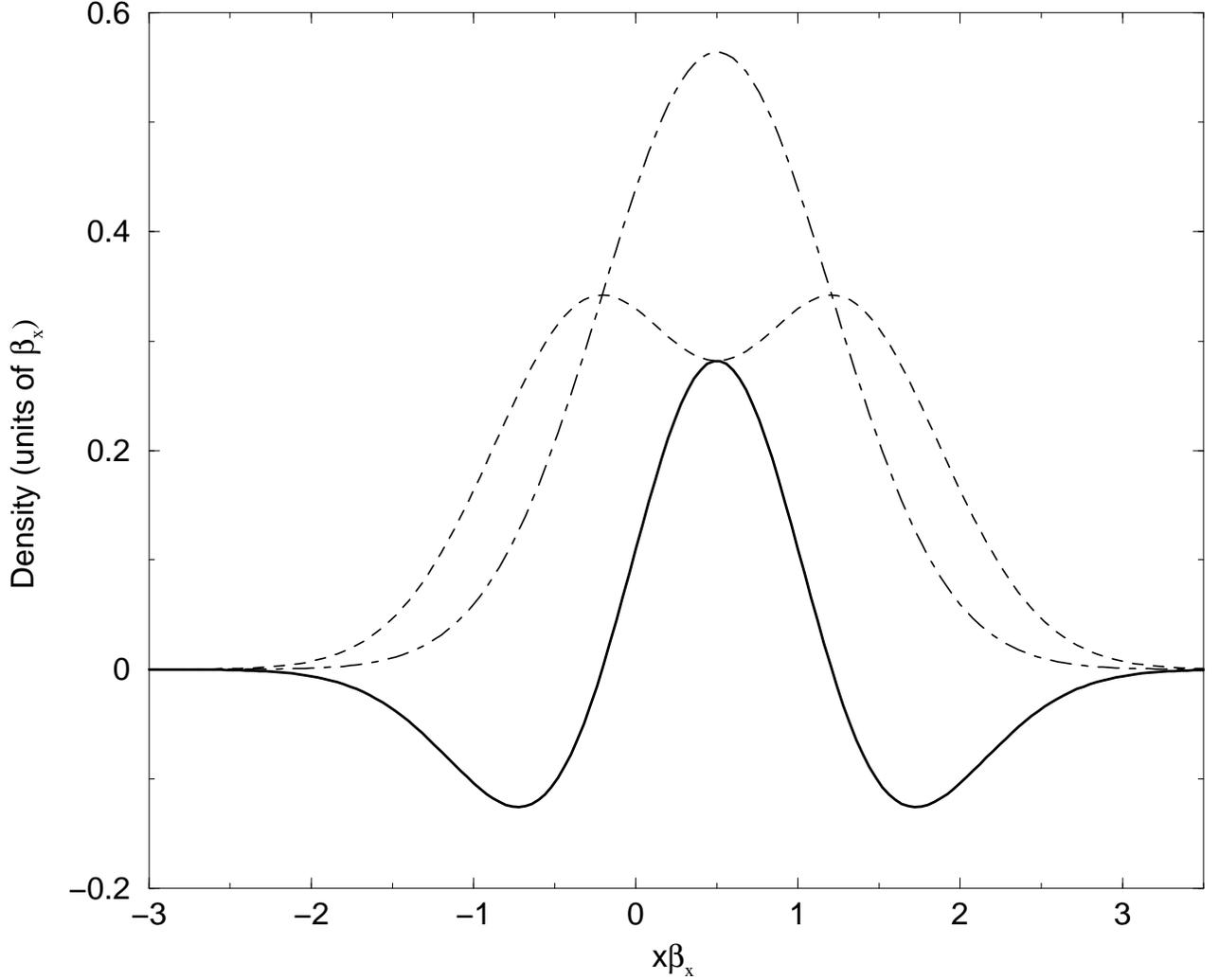} 
\caption{Changes of electron density at $E=E_c$, in the longitudinal direction (parallel to the applied field) of free-electrons trapped in a harmonic potential with an additional weak anharmonic term, $\lambda x^4$. In our example, the ground state 
configuration, $\prod n_y^{A_{n_y}}$, changes from $0^4$ to 
$0^21^2$, at $E=E_c$. The dashed lines show the longitudinal density at a value of the field slightly smaller than $E_c$. The dotted-dashed lines show the longitudinal density at a value of the field slightly higher than $E_c$. The full lines show the changes in the longitudinal density at the transition: electrons are transfered from the edges to the center of the dot. Here, we take $\beta_xx_E=0.5$ (see text).} 
\end{figure} 
 
\begin{figure} 
\includegraphics[width=14cm,angle=0]{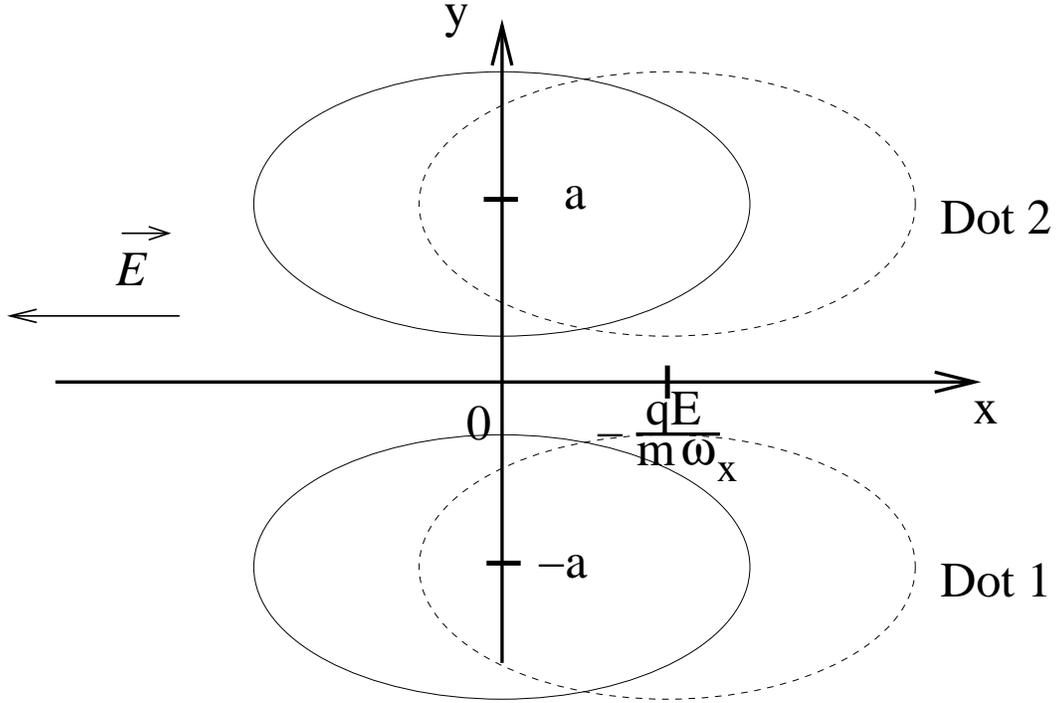} 
\caption{Two laterally coupled rectangular quantum dots with a harmonic confinement such that $\omega_x<\omega_y$ (full lines). An applied electric field along the x-axis shifts the position of the center of mass of the two quantum dots, supposed to be identical, by $x_E=-\frac{qE}{m\omega_x}$ (dashed lines).} 
\end{figure} 
 
\begin{figure} 
\includegraphics[width=14cm,angle=0]{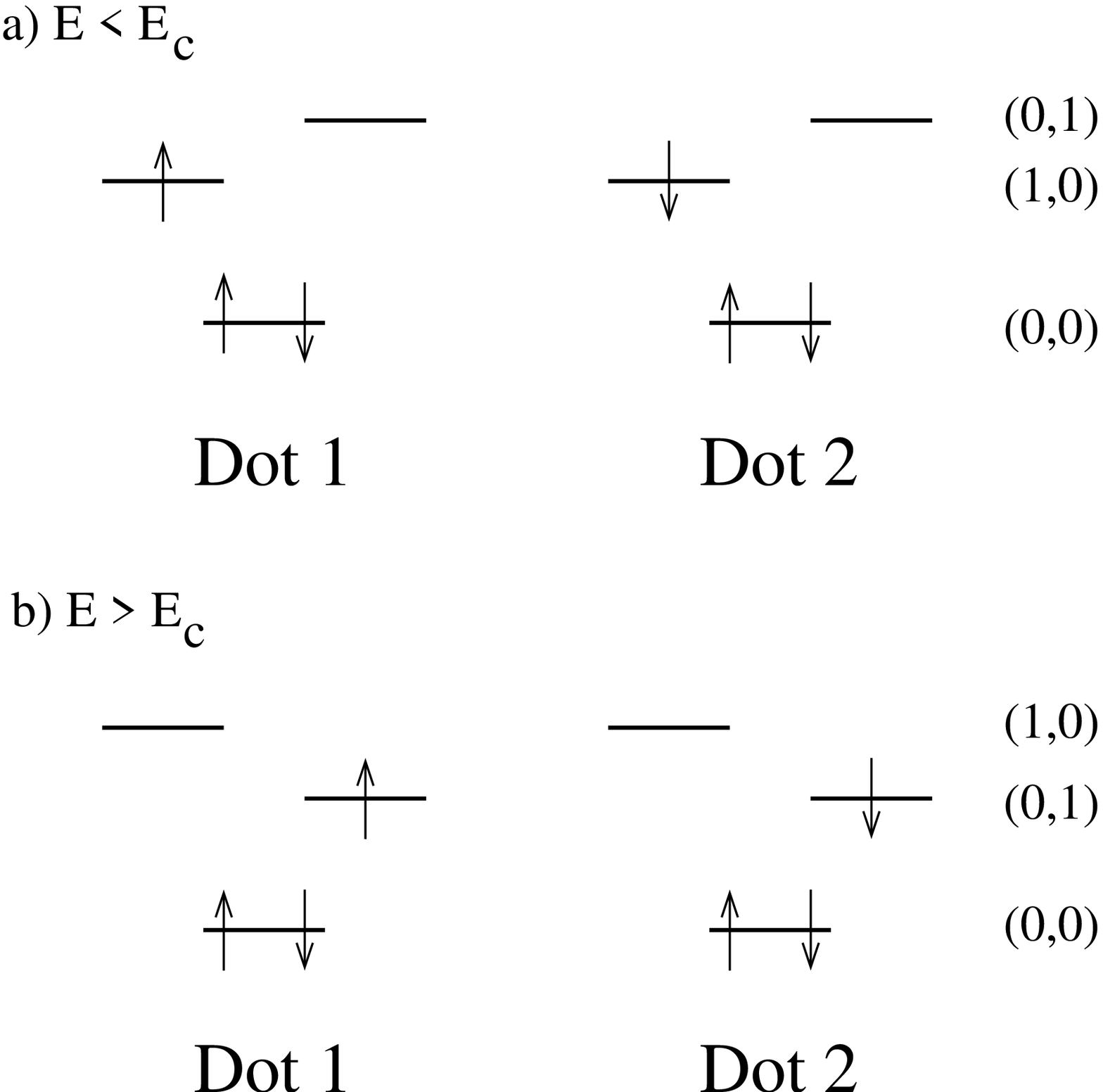} 
\caption{Example of electronic configuration of two weakly-coupled quantum dots with three electrons in each. An applied electric field can change the energy order of the one-electron states, $(n_x,n_y)$, in each dot. a) For $E<E_c$, the states $(0,0)$ and $(1,0)$ are populated. b) For $E>E_c$, the states $(0,0)$ and $(0,1)$ are populated. At the critical field, $E_c$, the unpaired electrons are transfered from the state $(1,0)$ to the state $(0,1)$ in each dot.} 
\end{figure} 
 
\begin{figure} 
\includegraphics[width=14cm,angle=0]{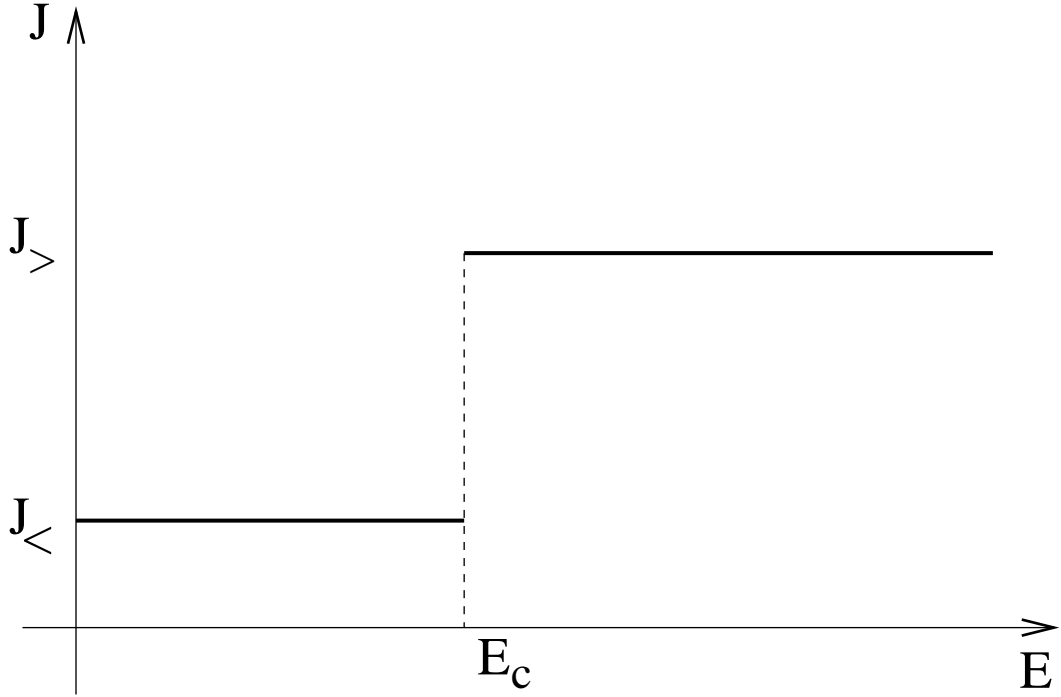} 
\caption{Exchange integral of the effective Heisenberg model (\ref{heisenberg}), describing the low-electronic properties of two weakly coupled quantum dots with one unpaired electron in each, as function of the applied electric field. At a critical field, $E_c$, the electronic structure of each dot is modified which causes a sudden jump of $J$, changing from $J^<$ to $J^>$. This property may be used to perform a {\it swap} operation (see text) in the context of quantum computation. } 
\end{figure} 
 
\end{document}